\documentclass[10pt,letterpaper]{article}
\usepackage[top=0.85in,left=2.75in,footskip=0.75in]{geometry}
\usepackage{amsmath,amssymb}

\usepackage{changepage}

\usepackage[utf8x]{inputenc}
\usepackage[vietnamese, english]{babel}

\usepackage{textcomp,marvosym}

\usepackage{cite}

\usepackage{nameref,hyperref}

\usepackage{microtype}
\DisableLigatures[f]{encoding = *, family = * }

\usepackage[table]{xcolor}

\usepackage{array}

\usepackage{multirow}
\usepackage{listings}
\usepackage{color}
\usepackage{pythonhighlight}

\newcolumntype{+}{!{\vrule width 2pt}}

\newlength\savedwidth

\newcommand\thickhline{\noalign{\global\savedwidth\arrayrulewidth\global\arrayrulewidth 2pt}%
\hline
\noalign{\global\arrayrulewidth\savedwidth}}

\raggedright
\usepackage[aboveskip=1pt,labelfont=bf,labelsep=period,justification=raggedright,singlelinecheck=off]{caption}

\makeatletter
\renewcommand{\@biblabel}[1]{\quad#1.}
\makeatother

\usepackage{lastpage,fancyhdr,graphicx}
\usepackage{epstopdf}

\fancyheadoffset[L]{2.25in}
\fancyfootoffset[L]{2.25in}
\newcommand{\etal}{\textit{et al.} }

\begin{document}
\vspace*{0.2in}

\begin{flushleft}
{\Large
\textbf\newline{A clinical validation of VinDr-CXR, an AI system for detecting abnormal chest radiographs} 
}
\newline
\\
Ngoc Huy Nguyen\textsuperscript{1,\YinYang},
Ha Quy Nguyen\textsuperscript{2,3,\YinYang},
Nghia Trung Nguyen\textsuperscript{2},
Thang Viet Nguyen\textsuperscript{2},
Hieu Huy Pham\textsuperscript{2,3,*},
Tuan Ngoc-Minh Nguyen\textsuperscript{4}
\\
\bigskip
\textbf{1} Phu Tho Department of Health, Phu Tho, Vietnam
\\
\textbf{2} Medical Imaging Department, Vingroup Big Data Institute, Hanoi, Vietnam
\\
\textbf{3} College of Engineering and Computer Science, VinUniversity, Hanoi, Vietnam
\\
\textbf{4} Training and Direction of Healthcare Activities Center, Phu Tho General Hospital, Phu Tho, Vietnam
\\
\bigskip

%
%
\Yinyang These authors contributed equally to this work.





* Corresponding author. Email: v.hieuph4@vinbigdata.org
\end{flushleft}
\section*{Abstract}
Computer-Aided Diagnosis (CAD) systems for chest radiographs using artificial intelligence (AI) have recently shown a great potential as a second opinion for radiologists. The performances of such systems, however, were mostly evaluated on a fixed dataset in a retrospective manner and, thus, far from the real performances in clinical practice. In this work, we demonstrate a mechanism for validating an AI-based system for detecting abnormalities on X-ray scans, VinDr-CXR, at the Phu Tho General Hospital--a provincial hospital in the North of Vietnam. The AI system was directly integrated into the Picture Archiving and Communication System (PACS) of the hospital after being trained on a fixed annotated dataset from other sources. The performance of the system was prospectively measured by matching and comparing the AI results with the radiology reports of 6,285 chest X-ray examinations extracted from the Hospital Information System (HIS) over the last two months of 2020. The normal/abnormal status of a radiology report was determined by a set of rules and served as the ground truth. Our system achieves an F1 score--the harmonic average of the recall and the precision--of 0.653 (95\% CI 0.635, 0.671) for detecting any abnormalities on chest X-rays. Despite a significant drop from the in-lab performance, this result establishes a high level of confidence in applying such a system in real-life situations.



\section*{Introduction}
Chest radiograph, or chest X-ray (CXR), remains one of the most common, yet hard to interpret, imaging protocols in medicine. It is hoped that a Computer-Aided Diagnosis (CAD) system using artificial intelligence (AI) can effectively assist radiologists and help mitigate the misdiagnosis rate on CXRs. Leveraging recent advances in deep learning~\cite{Zhou_2021}, such systems have achieved a great success in detecting a wide range of abnormalities on CXRs~\cite{rajpurkar2017chexnet,rajpurkar2018deep,irvin2019chexpert,majkowska2020chest,rajpurkar2020chexpedition,tang2020automated,pham2020interpreting,hwang2019development,lakhani2017deep,pasa2019efficient,Liu2020Tuberculosis}. Most of the existing systems are supervised-learning models that were trained and validated on different parts of a dataset that was collected and labeled in a retrospective fashion.  For example, several deep learning models were developed~\cite{rajpurkar2018deep,majkowska2020chest} on the ChestX-ray14 dataset~\cite{wang2017chest14} for classifying 14 common thoracic pathologies.  Recently, most algorithms for detecting abnormalities on CXRs were trained and validated on the CheXpert~\cite{irvin2019chexpert,pham2020interpreting,rajpurkar2020chexpedition} and MIMIC-CXR~\cite{Johnson2019mimic} datasets, which include the same set of 14 findings that are slightly different from the labels of ChestX-ray14. The performances of the aforementioned AI systems in differentiating multiple findings on CXRs were reported to be comparable with radiologists. Other works were devoted to detecting a specific lung disease such as pneumonia~\cite{rajpurkar2017chexnet}, pulmonary tuberculosis~\cite{lakhani2017deep,Liu2020Tuberculosis} and lung cancer~\cite{Ausawalaithong2018cancer}. Notably, Rajpurkar \etal \cite{rajpurkar2017chexnet} trained a convolutional neural network (CNN) for detecting pneumonia that achieved an F1 score of  0.435 (95\% CI 0.387, 0.481) on the ChestX-ray14 dataset, which performance was claimed to exceed those of practicing radiologists. Tang \etal proposed~\cite{Tang2020abnormal} to train an abnormality classifier, which is closely related to our work, with various CNN architectures over 3 CXR different datasets: the ChestX-ray14, the RSNA Pneumonia Detection Challenge~\cite{rsna2018pneumonia} and the Indiana University Hospital Network~\cite{Fushman2016Indiana}. Although reaching impressive AUC (Area under receiver operating characteristic Curve) performances of 0.9x, those models were again evaluated on retrospective curated datasets that might be drastically different from the real data in clinical settings.

Unlike existing works, our study does not focus on the development and the retrospective evaluation of an AI-based CAD system for CXR. Instead, we propose a framework to validate such a system while being deployed at a clinical site for a significantly long period. In particular, we integrate our system, VinDr-CXR, directly to the Picture Archiving and Communication System (PACS) of the Phu Tho General Hospital--a provincial hospital in Vietnam. The systems consists of 3 AI models that were trained on our own dataset~\cite{nguyen2021vindrcxr} collected from other sources. All CXRs generated by the PACS during two months are prospectively automatically analyzed by the VinDr-CXR. The obtained AI results are then matched and compared with the radiology reports that are extracted from the Hospital Information System (HIS) to compute the performance of the system in distinguishing abnormal versus normal CXR studies. Despite the ability of the system to localize multiple classes of lesions, we only measure its performance as a binary classifier. The reason for doing so is that it is much more reliable to decide if a radiology report is abnormal than to interpret its subtle details. We also propose simple template matching rules to determine the normal/abnormal status of a report.

Over the last two months of 2020, the VinDr-CXR system generated AI results for 6,687 CXR studies taken at the Phu Tho General Hospital, 6,285 of which were matched with corresponding radiology reports from the HIS. The matching was nontrivial since the PACS and the HIS were not linked by accession numbers. Instead, we had to rely on the patient ID and other attributes of the DICOM files and the radiology reports. By treating the normal/abnormal status of the 6,285 matched reports as a ground-truth reference, the abnormality classifier of the VinDr-CXR yielded an F1 score of 0.653 (95\% Confidence Interval (CI) 0.635, 0.671). The 95\% CI of the F1 score statistic was obtained by bootstrapping~\cite{EfroTibs93}, a method that was also used in~\cite{rajpurkar2017chexnet}. The F1 score obtained in this clinical setting is significantly below the one achieved while training and validating the model ``at home'' on a retrospective dataset. Nonetheless, the reported performance still gives us a high level of confidence in deploying the VinDr-CXR system in clinical practice. It also serves a good baseline for similar AI-based CAD systems to be clinically validated.  

\section*{Materials and methods}
\begin{figure}[!h]
	\centering
	\includegraphics*[width=0.8\linewidth]{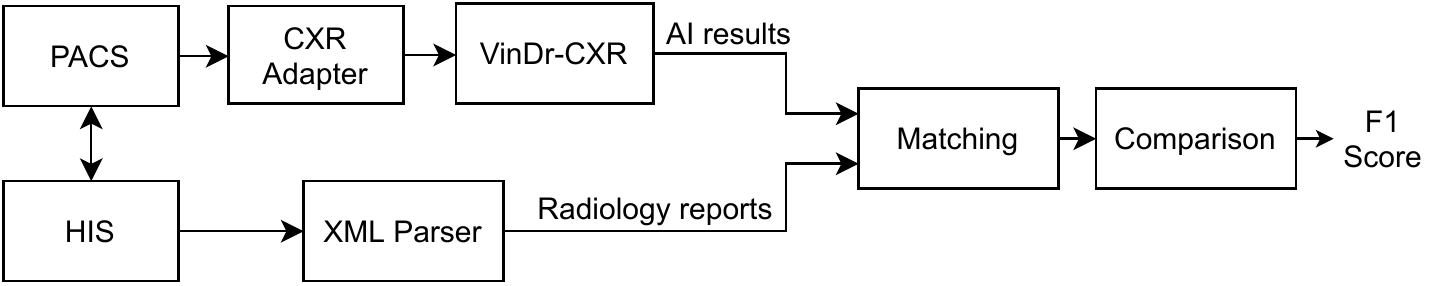}
	\caption{{\bf Validation scheme.}
		PACS and HIS are linked by Patient ID.}
	\label{fig:validation_diagram}
\end{figure}

We propose an overall scheme for validating our CAD system, VinDr-CXR, as illustrated in Fig~\ref{fig:validation_diagram}. A set of AI results are obtained by directly integrating VinDr-CXR into the PACS of the Phu Tho General Hospital, while the corresponding radiology reports are extracted from the HIS via an XML parser. These two sets of results are then pairwise matched  and compared to each other to determine the correctness of the system in detecting abnormal CXRs. The final result of the validation will be reported as an F1 score, a metric that balances the precision and recall of a binary classifier. 

\subsection*{Development of AI models}
\begin{figure}[!h]
	\centering
	\includegraphics*[width=\linewidth]{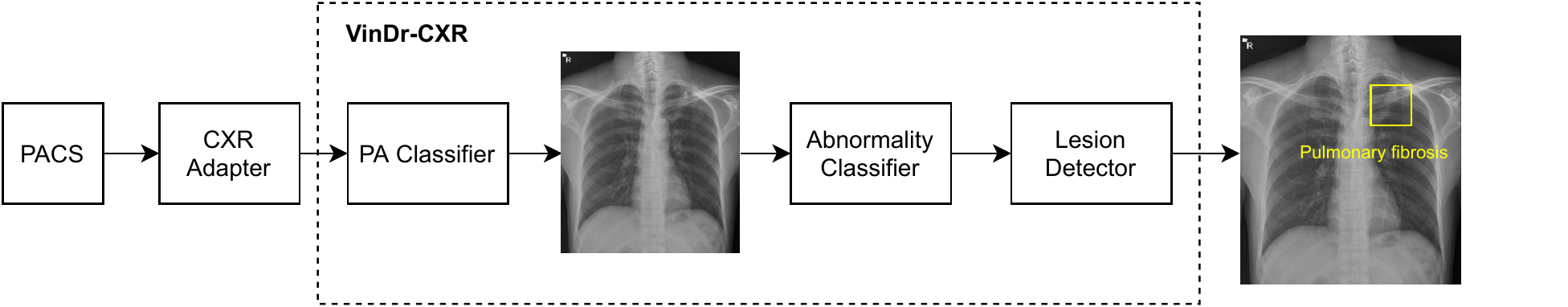}
	\caption{{\bf VinDr-CXR pipeline.}
		The system includes 3 concatenated AI models that are integrated  to the PACS via a CXR adapter. The output of the system is the probability of the CXR being abnormal and the locations of the lesions, if any.}
	\label{fig:VinDr-CXR}
\end{figure}

As shown in Fig~\ref{fig:VinDr-CXR}, the VinDr-CXR system is a concatenation of 3 AI models: the PA classifier, the abnormality classifier, and, finally, the lesion detector. This system takes as input a CXR from the PACS and returns the probability that the image is abnormal and the locations of multiple classes of lesions, if any. All constitutional models of VinDr-CXR were obtained by training deep neural networks entirely on our own dataset, also called VinDr-CXR, part of which was made publicly available~\cite{nguyen2021vindrcxr}. This dataset was retrospectively collected from our partner hospitals in Vietnam and got annotated by a team of experienced radiologists. Each image in the dataset was manually labeled by at least 1 radiologist with a list of 6 different diagnoses where 22 types of lesions were annotated with bounding boxes. It is important to emphasize that none of the training data was  from the Phu Tho General Hospital. Each individual model was trained and validated before being deployed in the real clinical workflow of the hospital. We did not make any changes to the models during the two months of the clinical trial. This is to ensure that our models are not at all biased to the real-life validation setup. We briefly describe here the development of the 3 AI models; details of the training will be presented somewhere else. 
\subsubsection*{PA classifier}
This PA classifier is attached to the CXR adapter to guarantee that only CXRs of Posterior-Anterior (PA) view will be passed to the abnormality classifier, which was trained only on this type of images. The output of the PA classifier is a probability of the input image being a PA-view CXR. If this probability is greater than a normalized threshold of 0.5, the image will go through to the abnormality classifier; otherwise, the system will output an indicator that the image is invalid. The PA classifier adopted the ResNet-18 architecture~\cite{resnet} that was trained and validated on a dataset of total 9,864 scans where 4,329 of them are actual PA-view CXRs taken from the VinDr-CXR dataset. The negative training examples included lateral-view CXRs and images of other body parts that sometimes got through the CXR filter due to mismatched DICOM tags. The trained PA classifier achieves an F1 score of 0.980 on a validation set of 4,192 images. Here, the F1 score metric is defined as
\begin{equation}\label{eq:F1}
F_1 = \frac{TP}{TP + (FP+FN)/2},
\end{equation}
where $TP, FP, FN$ denote the numbers of true positive, false positive, and false negative samples, respectively.
\subsubsection*{Abnormality classifier}
The abnormality classifier separates abnormal CXRs from normal ones. It takes as input a PA-view CXR and outputs the probability that the image contains abnormal findings. Only images whose abnormal probabilities are above 0.5 will go to the lesion detector.  We trained the abnormality classifier as an EfficientNet-B6~\cite{efficientNet} on a dataset of 38,065 PA-view CXRs. All images that were labeled with ``No finding'' by the radiologists were treated as negative samples, while the rest are considered positive. This model was validated on another dataset of 9,611 images with an F1 score of 0.831. In this study, only the output of the abnormality classifier will be compared to the radiology reports to measure the performance of the whole system. 

\subsubsection*{Lesion detector}
The role of the lesion detector is to localize all findings on an abnormal CXR with bounding boxes and, at the same time, classify them into different types of lesion. That is, the system can tell not only whether a CXR is abnormal but also why it is and where the abnormalities come from. An example output of the lesion detector is visualized in Fig~\ref{fig:VinDr-CXR} where a bounding box of the class ``Pulmonary fibrosis'' is drawn around the lesion. Out of the 22 local classes in the VinDr-CXR dataset, we only trained the lesion detector on the 17 most prevalent ones as listed in Table~\ref{table1}. The training was performed on 23,524 abnormal CXRs with an EfficientDet-ED4 model~\cite{efficientDet}. The performance of the lesion detector was evaluated on a validation set of 4,470 images using the Average Precision (AP) metric at the Intersection-over-Union (IoU) threshold of 0.4 or shortly AP\,@\,0.4. This is a standard metric for objection detection models in computer vision~\cite{pascalVOC,coco}. Table 1 reports the performance of the lesion detector in terms of AP for each of the 17 classes with an average AP (mAP) of 0.365. However, the model was not validated on the data the Phu Tho General Hospital. This is due to the hardness in interpreting the lesion locations in a radiology report. 
\begin{table}[!ht]
	\centering
	\caption{
		{\bf Performance of the lesion detector on 17 classes of abnormality.} }
	\begin{tabular}{|l|l|}
		\hline
		\textbf{Lesion class}  & \textbf{AP\,@\,0.4} \\ \hline
		Aortic enlargement              & 0.663               \\ \hline
		Atelectasis                     & 0.231               \\ \hline 
		Calcification                   & 0.272               \\ \hline 
		Cardiomegaly                    & 0.860               \\ \hline
		Clavicle fracture               & 0.459               \\ \hline 
		Consolidation                   & 0.281               \\ \hline
		Emphysema                       & 0.185               \\ \hline
		Enlarged PA                     & 0.256               \\ \hline 
		Infiltration                    & 0.318                \\ \hline
		Interstitial lung disease (ILD) & 0.315               \\ \hline 
		Nodule/Mass                     & 0.251               \\ \hline
		Opacity                         & 0.197               \\ \hline 
		Pleural effusion                & 0.387               \\ \hline 
		Pleural thickening              & 0.228               \\ \hline 
		Pneumothorax                    & 0.579               \\ \hline
		Pulmonary fibrosis              & 0.340               \\ \hline
		Rib fracture                    & 0.381               \\ \thickhline
		\textbf{mAP}                    & \textbf{0.365}      \\ \hline
	\end{tabular}
	\label{table1}
\end{table}
\subsection*{Integration of AI models to PACS}
As shown in Fig~\ref{fig:VinDr-CXR}, the VinDr-CXR system is integrated to the PACS of the hospital through the CXR adapter. This is a web service that pulls all images from the PACS via the DICOMweb protocol~\cite{DICOMweb} and only passes CXRs to the AI models of the VinDr-CXR. To check if a scan is a CXR, we rely on the  
\verb|MODALITY| and the \verb|BODY_PART_EXAMINIED| attributes of the DICOM file. In particular, the AI models are triggered only if the value of \verb|MODALITY| is either \verb|CR|, \verb|DR| or \verb|DX| and the value of \verb|BODY_PART_EXAMINIED| is either \verb|CHEST| or \verb|THORAX|. These conditions were established by surveying the imaging procedure at the Radiology Department of the Phu Tho General Hospital. For deploying the VinDr-CXR at other clinical sites, the CXR adapter might be slightly modified to catch all CXRs from the PACS.  
\subsection*{Extraction of radiology reports from HIS}
\begin{figure}[!h]
	\centering
	\includegraphics*[width=\linewidth]{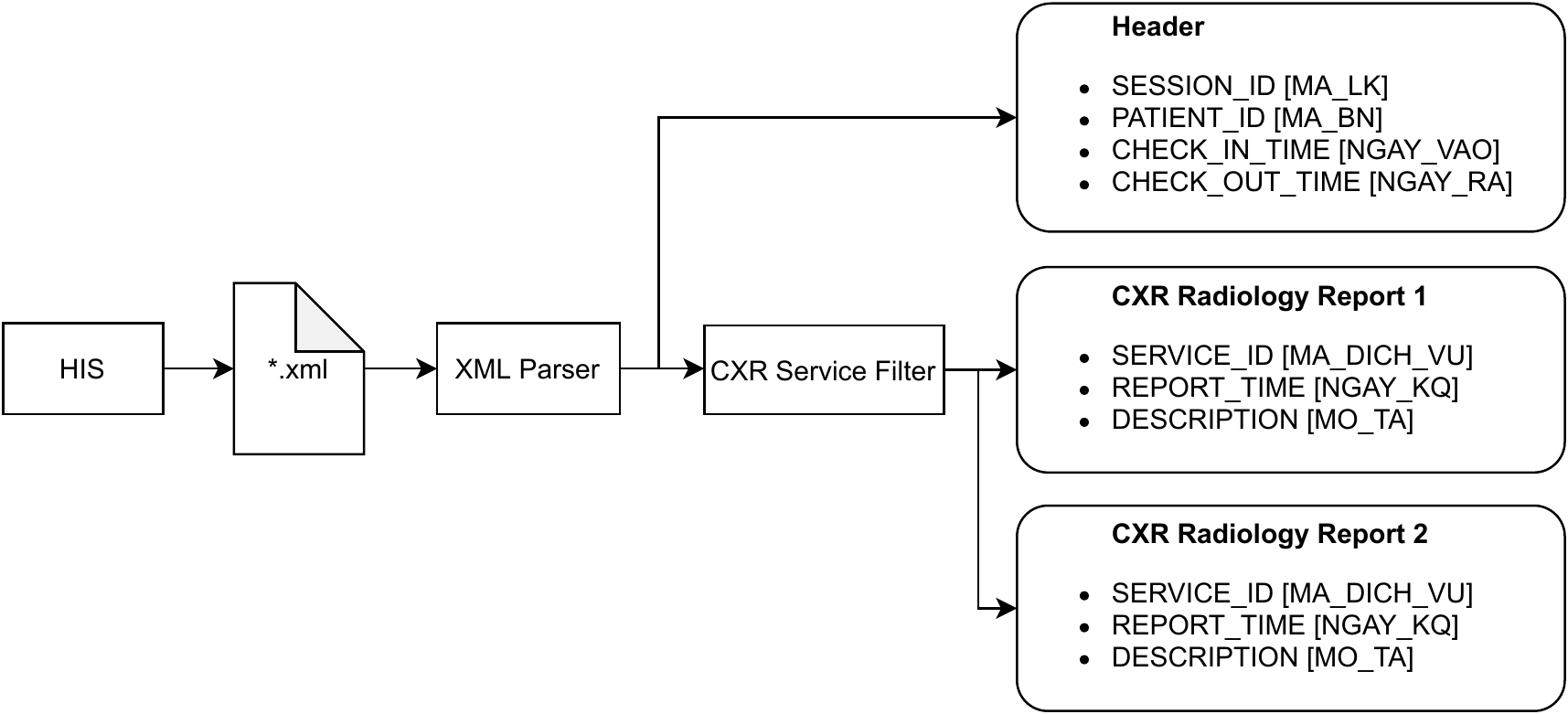}
	\caption{{\bf Procedure for extracting all radiology reports of CXR examinations from HIS.} The original names of the attributes, which are in Vietnammese, are put inside square brackets.}
	\label{fig:XML_parser}
\end{figure}

The radiology reports have to be extracted from HIS according to a procedure described in Fig~\ref{fig:XML_parser}. Each session of examination and treatment is stored in a single Extensible Markup Language (XML) file that can be exported from HIS. A session includes all information of the patient from the check-in time to the check-out time. The XML parser is used to read all the reports within a session, each of which includes the \verb|SERVICE_ID|, \verb|REPORT_TIME|, and \verb|DESCRIPTION| attributes. The CXR service filter only keeps the reports whose \verb|SERVICE_ID| matches a fixed value reserved for the CXR imaging by the Vietnamese Ministry of Health. The XML parser can also read the header of a session that includes \verb|SESSION_ID|, \verb|PATIENT_ID|, \verb|CHECK_IN_TIME|, and \verb|CHECK_OUT_TIME|. These attributes are shared among all radiology reports within the session and will be used, in addition to the \verb|REPORT_TIME|, to match a radiology report with an AI result of a CXR scan.   

\subsection*{Matching AI results with radiology reports}
\begin{figure}[!h]
	\centering
	\includegraphics*[width=0.8\linewidth]{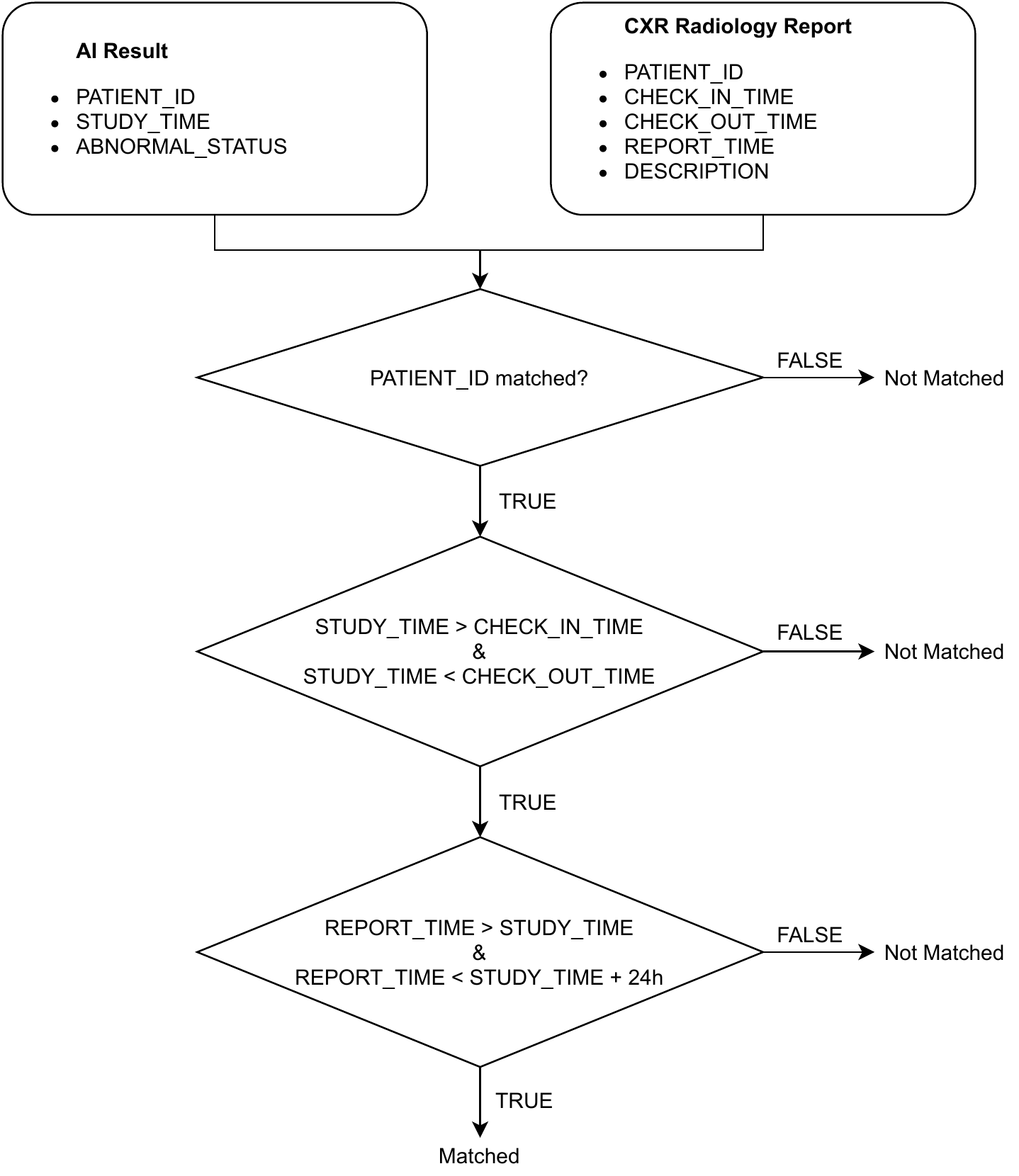}
	\caption{{\bf Algorithm for matching an AI result with a radiology report.} }
	\label{fig:matching_algo}
\end{figure}

After extracting the CXR radiology reports from HIS, we match each of them with an AI result obtained from VinDr-CXR via the algorithm illustrated in Fig~\ref{fig:matching_algo}. Note that an AI result includes the \verb|ABNORMAL_STATUS| (0/1) of a CXR study and is associated with a \verb|PATIENT_ID| and a \verb|STUDY_TIME|, which attributes are extracted from the DICOM file. Since HIS and PACS are linked by the \verb|PATIENT_ID|, the matching algorithm uses this key to check whether an AI result and a radiology report are of the same patient. Next, the \verb|STUDY_TIME| has to be in between the \verb|CHECK_IN_TIME| and the \verb|CHECK_OUT_TIME|. Finally, the \verb|REPORT_TIME| must be within 24 hours from the \verb|STUDY_TIME|, which is a regulated protocol of the hospital. If all the aforementioned conditions are satisfied, the AI result and the CXR radiology report are matched.

\subsection*{Comparing AI results to radiology reports}
The \verb|ABNORMAL_STATUS| of an AI result is then compared to the \verb|DESCRIPTION| of the matched radiology report, if any, to measure the performance of VinDr-CXR in detecting abnormal CXR scans. To that end, we propose a simple template-matching rule to determine if a description is normal. In particular, we observe that a CXR radiology report without any findings always includes 4 paragraphs that describe the 4 fixed anatomical regions of the thorax: chest wall, pleura, lung, and mediastinum. The templates for normal descriptions of these 4 regions are summarized in Table~\ref{table2}. A region is considered normal if one of the corresponding templates exactly appears in the \verb|DESCRIPTION| of a radiology report. A report is normal if all the 4 regions are normal, otherwise it is abnormal. 
\begin{table}[!ht]
	\centering
	\caption{
		{\bf Templates for normal descriptions of the 4 anatomical regions in a CXR radiology report.} }
	\selectlanguage{vietnamese}
	\begin{tabular}{|l|l|}
		\hline
		\textbf{Anatomical region}   & \textbf{Templates for normal descriptions}     \\ \hline
		\multirow{2}{*}{Chest wall}  & \verb|không thấy hình bất thường xương lồng ngực|     \\ \cline{2-2} 
		& \verb|không thấy hình tổn thương xương lồng ngực|     \\ \thickhline 
		\multirow{3}{*}{Pleura}      & \verb|không thấy hình tràn dịch màng phổi|            \\ \cline{2-2} 
		& \verb|không thấy hình tràn dịch, khí màng phổi|       \\ \cline{2-2} 
		& \verb|không thấy hình tràn khí, tràn dịch màng phổi|  \\ \thickhline 
		Lung                         & \verb|nhu mô phổi không thấy bất thường|            \\ \thickhline 
		\multirow{2}{*}{Mediastinum} & \verb|hình tim và trung thất không thấy bất thường| \\ \cline{2-2} 
		& \verb|hình tim và trung thất bình thường|           \\ \hline
	\end{tabular}
	\selectlanguage{english}
	\label{table2}
\end{table}

\section*{Results and Discussion}
\begin{figure}[!h]
	\centering
	\includegraphics*[width=0.5\linewidth]{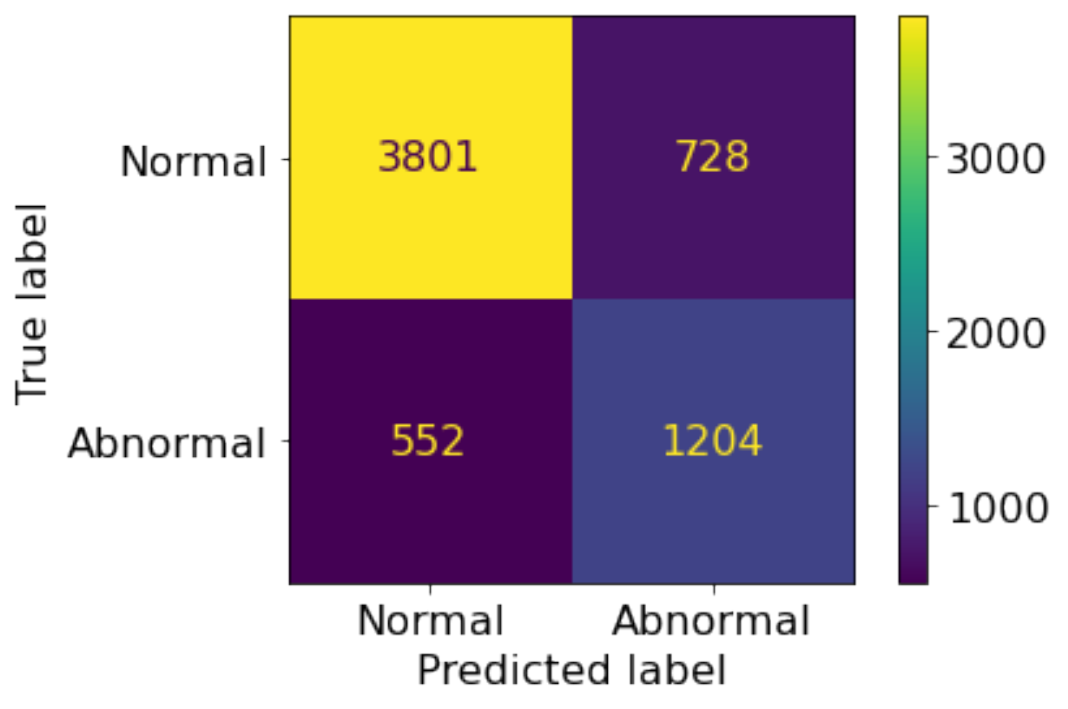}
	\caption{{\bf Confusion matrix of the VinDr-CXR abnormality classifier.} }
	\label{fig:confusion_matrix}
\end{figure}

\begin{figure}[!h]
	\centering
	\includegraphics*[width=0.7\linewidth]{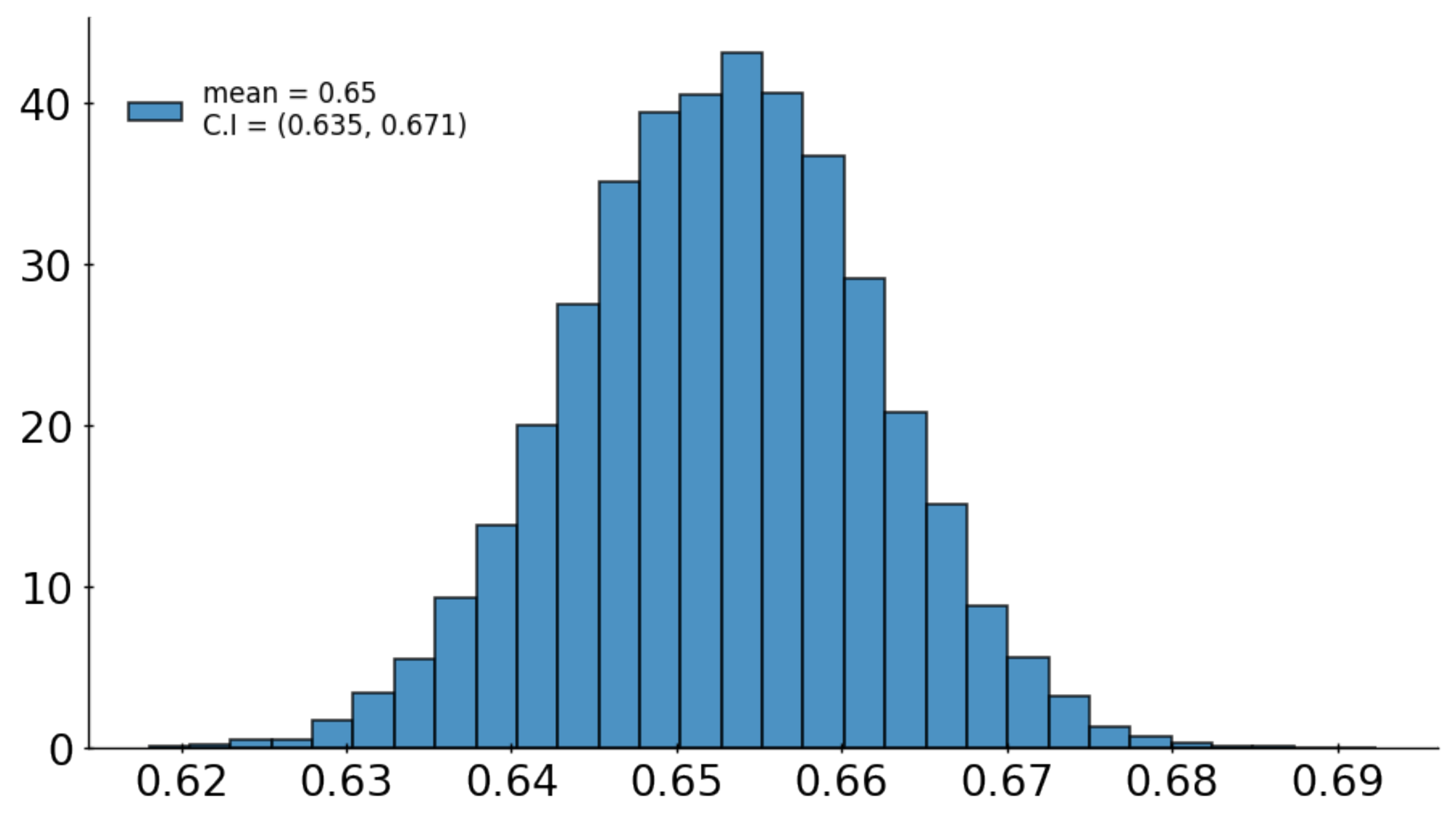}
	\caption{{\bf Bootstrap distribution of F1 scores of the VinDr-CXR abnormality classifier over 10,000 samples drawn from 6,285 studies.} }
	\label{fig:bootstrap}
\end{figure}

A set of 6,585 AI results was obtained by running the VinDr-CXR on all DICOM images in the PACS of the Phu Tho General Hospital throughout November and December of 2020. Meanwhile, another set of 6,687 CXR radiology reports was extracted from the HIS during this period. Applying the matching algorithm to these two sets resulted in 6,285 studies of 5,989 patients that have both an AI result and a radiology report. By matching the radiology reports of these studies with the templates given in Table~\ref{table2}, we achieved a ground truth of 4,529 (72.4\%) normal  and 1,756 (27.6\%) abnormal cases. The confusion matrix of the VinDr-CXR abnormality classifier over the total 6,285 studies is plotted in Fig~\ref{fig:confusion_matrix}.  We follow~\cite{rajpurkar2017chexnet} to compute the average F1 score on 10,000 bootstrap samples drawn with replacement~\cite{EfroTibs93} from the 6,285 studies. We also use the 2.5th and 97.5th percentiles of the bootstrap distribution to establish the 95\% confidence interval (CI). The bootstrap distribution of F1 scores is shown in Fig~\ref{fig:bootstrap}, which gives a mean F1 score of 0.653 (95\% CI 0.635, 0.671).

It can be seen that the F1 score of our AI system in detecting abnormal CXRs significantly drops from 0.831 to 0.653 when shifting from the training settings to the deployment phase in a clinical site. This might be caused by the shift in the distribution of the CXR images or the additional clinical information received by the radiologists in practice. Furthermore, it must be admitted that the radiologists' reports at the deployed site are not necessarily all accurate. Nevertheless, the obtained F1 score still exhibits a high level of confidence when deploying the VinDr-CXR in practice, as compared to the 0.435 F1 score of the pneumonia detector presented in~\cite{rajpurkar2017chexnet}.  

One limitation of this study is that it only provides a coarse evaluation of the VinDr-CXR system for the task of classifying a CXR into normal and abnormal categories. To validate the lesion detector of the system, we need a more sophisticated interpretation of the radiology reports that can extract a ground truth comparable to the output of the AI model. Another shortcoming of this work is that the reported clinical validation is for the AI system itself. It is even more important to assess the effect of such a system on improving the quality of the radiologists' diagnoses. Moreover, an ideal AI-based CAD system should continuously learn from the daily feedback of the doctors rather than staying stationary like our VinDr-CXR. We plan to address all of these drawbacks in our future research agenda.

\section*{Conclusion}
We have discussed in this paper a mechanism for validating the performance of the VinDr-CXR system in classifying normal/abnormal chest radiographs at the Phu Tho General Hospital during the last two months of 2020. Once the AI models of the system were trained on an annotated dataset from different sources, they were directly integrated into the PACS of the hospital and never got retrained during the validation period. The performance of the abnormality classifier was prospectively measured by matching and then comparing the obtained AI results with a set of radiology reports exported from the HIS. Since the PACS and the HIS were linked only by patient IDs, we proposed an algorithm to match an AI result of a study with a CXR radiology reports. We also adopted a simple template matching rule to decide the abnormal status of a radiology report, which served as a ground-truth reference. We obtained an average F1 score of 0.653 (95\% CI 0.635, 0.671) for the abnormality classifier over 10,000 resamples drawn from the 6,285 studies of 5,989. We believe this result has set a significant benchmark for deploying AI systems for chest radiograph analysis in clinical practice. 

\section*{Acknowledgments}
This work was funded by the Vingroup JSC.


%
%
%
%
%
%
%

\bibliography{mylib}

\end{document}